\documentclass[twocolumn]{aastex631} %linenumbers
\hypersetup{linkcolor=red,citecolor=blue,filecolor=cyan, urlcolor=black}
\usepackage{multirow}
\usepackage{amsmath}
\usepackage{amssymb}
\usepackage{balance}
\received{March 30, 2022}
\revised{April 7, 2022}
\accepted{April 8, 2022}
\submitjournal{ApJS}

\shorttitle{66,000 Mira stars in the Milky Way}
\shortauthors{Iwanek et al.}

\begin{document}

\title{The OGLE Collection of Variable Stars. Nearly 66,000 Mira stars in the Milky Way}

\correspondingauthor{Patryk Iwanek}
\email{piwanek@astrouw.edu.pl}

\author[0000-0002-6212-7221]{Patryk Iwanek}
\affiliation{Astronomical Observatory, University of Warsaw, Al. Ujazdowskie 4, 00-478 Warsaw, Poland}

\author[0000-0002-7777-0842]{Igor Soszy{\'n}ski}
\affiliation{Astronomical Observatory, University of Warsaw, Al. Ujazdowskie 4, 00-478 Warsaw, Poland}

\author[0000-0003-4084-880X]{Szymon Koz{\l}owski}
\affiliation{Astronomical Observatory, University of Warsaw, Al. Ujazdowskie 4, 00-478 Warsaw, Poland}

\author[0000-0002-9245-6368]{Radosław Poleski}
\affiliation{Astronomical Observatory, University of Warsaw, Al. Ujazdowskie 4, 00-478 Warsaw, Poland}

\author[0000-0002-2339-5899]{Paweł Pietrukowicz}
\affiliation{Astronomical Observatory, University of Warsaw, Al. Ujazdowskie 4, 00-478 Warsaw, Poland}

\author[0000-0002-2335-1730]{Jan Skowron}
\affiliation{Astronomical Observatory, University of Warsaw, Al. Ujazdowskie 4, 00-478 Warsaw, Poland}

\author[0000-0002-3051-274X]{Marcin Wrona}
\affiliation{Astronomical Observatory, University of Warsaw, Al. Ujazdowskie 4, 00-478 Warsaw, Poland}

\author[0000-0001-7016-1692]{Przemysław Mróz}
\affiliation{Astronomical Observatory, University of Warsaw, Al. Ujazdowskie 4, 00-478 Warsaw, Poland}

\author[0000-0001-5207-5619]{Andrzej Udalski}
\affiliation{Astronomical Observatory, University of Warsaw, Al. Ujazdowskie 4, 00-478 Warsaw, Poland}

\author[0000-0002-0548-8995]{Michał K. Szymański}
\affiliation{Astronomical Observatory, University of Warsaw, Al. Ujazdowskie 4, 00-478 Warsaw, Poland}

\author[0000-0001-9439-604X]{Dorota M. Skowron}
\affiliation{Astronomical Observatory, University of Warsaw, Al. Ujazdowskie 4, 00-478 Warsaw, Poland}

\author[0000-0001-6364-408X]{Krzysztof Ulaczyk}
\affiliation{Department of Physics, University of Warwick, Coventry CV4 7 AL, UK}
\affiliation{Astronomical Observatory, University of Warsaw, Al. Ujazdowskie 4, 00-478 Warsaw, Poland}

\author[0000-0002-1650-1518]{Mariusz Gromadzki}
\affiliation{Astronomical Observatory, University of Warsaw, Al. Ujazdowskie 4, 00-478 Warsaw, Poland}

\author[0000-0002-9326-9329]{Krzysztof Rybicki}
\affiliation{Astronomical Observatory, University of Warsaw, Al. Ujazdowskie 4, 00-478 Warsaw, Poland}

\author[0000-0002-3218-2684]{Milena Ratajczak}
\affiliation{Astronomical Observatory, University of Warsaw, Al. Ujazdowskie 4, 00-478 Warsaw, Poland}

\begin{abstract}

\noindent We present a collection of 65,981 Mira-type variable stars found in the Optical Gravitational Lensing Experiment (OGLE) project database. Two-thirds of our sample (40,356 objects) are located in the Galactic bulge fields, whereas 25,625 stars are in the Galactic disk. The vast majority of the collection (47,532 objects) are new discoveries. We provide basic observational parameters of the Mira variables: equatorial coordinates, pulsation periods, {\it I}-band and {\it V}-band mean magnitudes, \mbox{{\it I}-band} brightness amplitudes, and identifications in other catalogs of variable stars. We also provide the {\mbox{{\it I}-band}} and {\mbox{{\it V}-band}} time-series photometry collected since 1997 during the OGLE-II, OGLE-III, and OGLE-IV phases. The classical selection process, i.e., mostly based on the visual inspection of light curves by experienced astronomers, led us to the high purity of the catalog. As a result, this collection can be used as a training set in the machine learning classification algorithms. Using overlapping parts of adjacent OGLE fields, we estimate the completeness of the catalog to be about 96\%. We compare and discuss the statistical features of Miras located in different regions of the Milky Way. We show examples of stars that change their type over time, from a semi-regular variable to Mira and vice versa. This dataset is perfectly suited to study the three-dimensional structure of the Milky Way, and it may help to explain the puzzle of the X-shaped bulge.

\end{abstract}


\section{Introduction} \label{sec:introduction}

The Mira star ($o$ Ceti), named by Johannes Hevelius in 1662, in Latin means {\it wonderful, astonishing, miracle}. This name is not accidental, as in the 16th century, just after the discovery of Mira's variability by David Fabricius and its periodicity by Johannes Holwarda, the regular appearance and disappearance of the celestial object have never been seen before. The Mira star has become a prototype of a numerous group of pulsating long-period variables \citep[LPVs;][]{1933JRASC..27...75H}.

LPVs are mostly red giant branch (RGB) or asymptotic giant branch (AGB) stars, however, late-type pulsating supergiants (semi-regular variables of c type; SRc) are also part of this group. During these evolutionary stages stars are pulsationally unstable, which leads to their brightness changes with periods ranging from days to years. One of the subgroups of LPVs is Mira-type variables, which are perhaps the best-known LPVs due to the over four-century-long history of their studies.

The first discovered Mira-type star, after $o$ Ceti itself, was $\chi$ Cygni discovered by Gottfried Kirch in 1686. Between 1686 and 1796 four Miras were known, and the number of known Mira variables increased to 251 between 1796 and 1896 \citep{1997JAVSO..25..115H}. At the beginning of the 20th century, more and more Mira-type variables were discovered and their pulsation periods measured \citep[e.g. ][]{1928BAN.....4..174H, 1930JRASC..24..271L, 1957PZ.....12...33K, 1957PZ.....12..108K, 1960AJ.....65..381W}. After half a century of observations, members of the American Association of Variable Star Observers (AAVSO) collected three million measurements of visual magnitudes of Mira-type variables \citep{1955slpv.book.....C}. Along with the technological development, regular observations of Miras in the near-infrared (NIR) passbands began \citep{1971ApJ...169...63L, 1973ApJS...25..369B, 1975JAVSO...4...22W, 1976MNRAS.174..169L, 1979SAAOC...1...61C, 1982MNRAS.199..245G}. At the same time, intensive spectroscopic studies of Mira-type variables were carried out \citep{1966ApJS...13..333K, 1974ApJS...28..271K, 1982ApJ...252..697H}. \citet{1976A&A....48...27M, 1977ApJ...211..499W} and \citet{1979ApJ...227..220W} studied Miras' mass-loss phenomenon, which is still the subject of intense research \citep[e.g.][]{2019A&A...622A.120U, 2021arXiv210910730N}. The NIR period-luminosity relation (PLR) for Miras was first shown by \citet{1981Natur.291..303G}, based on 11 Miras from the Large Magellanic Cloud (LMC).

Mira-type variables could be identified by four fundamental observational properties: their pulsation period, brightness variation amplitude, color index, and shape of the light curve. Miras are fundamental-mode pulsators with periods ranging from about 80 days to over 1000 days. Recently, \citet{2022A&A...658L...1T} presented the analysis of the period-age relation for LPVs, including Miras. The authors showed that the pulsation period (of the fundamental mode; in general, other LPVs can pulsate in more than one mode) decreases with increasing age because of the dominant role of mass in shaping the stellar structure and evolution.

\begin{figure*}
\centering
\includegraphics[scale=0.37]{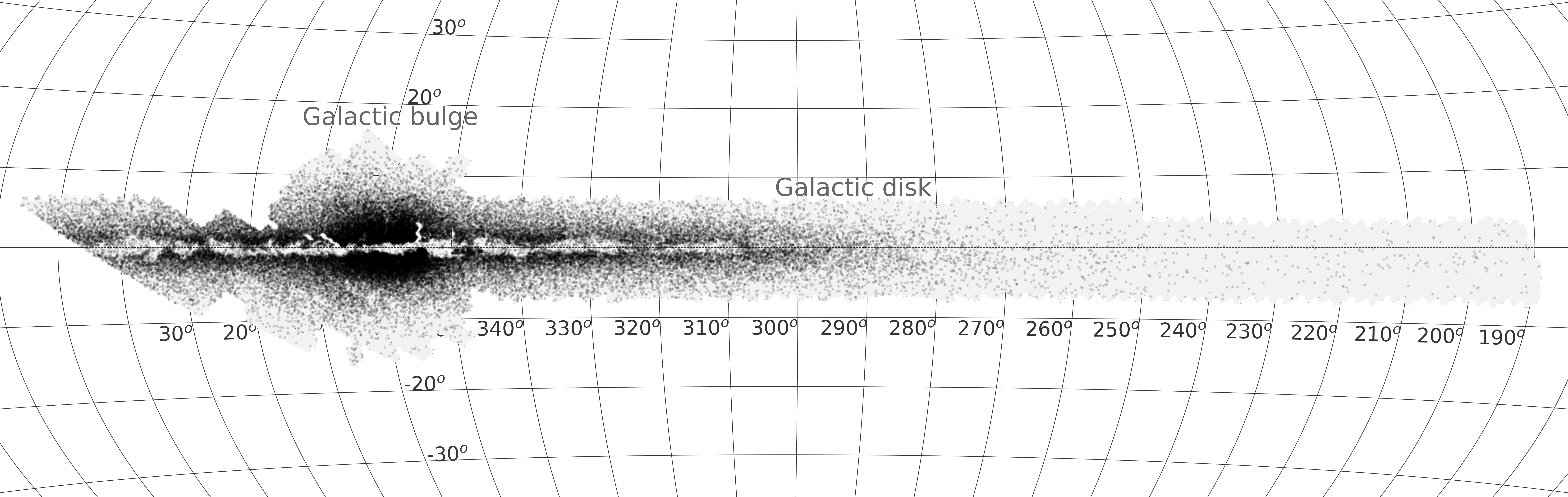}
\caption{Distribution of Miras (black points) in the Galactic coordinates. Our catalog contains 40,356 Mira-type variables in the Galactic bulge fields and 25,625 in the Galactic disk. The gray area shows the OGLE-IV footprint of the Milky Way.}
\label{fig:miras_map}
\end{figure*}

Miras show large brightness variations, which is the largest at short wavelengths \citep[the peak-to-peak amplitudes in the $V-$band are higher than \mbox{2.5 mag};][]{2017ARep...61...80S}, and decreases toward the infrared part of the light spectrum \citep[above 0.4 mag in the $K-$band;][]{2006MNRAS.369..751W, 2021MNRAS.500...82I, 2021ApJS..257...23I}. Mira-type variables with the lowest brightness amplitude smoothly turn into another LPVs subgroup -- semi-regular variables (SRVs). Miras and SRVs are customarily separated by amplitude, and stars with an {\it I}-band amplitude greater than 0.8 mag are classified as Mira-type \citep{2009AcA....59..239S, 2013AcA....63...21S}. This boundary is conventional, and misclassifications are possible around  the \mbox{$I-$band amplitude of $ 0.8$ mag}. Miras (and other LPVs) are divided into Oxygen-rich (O-rich) and Carbon-rich (C-rich) classes, depending on their surface composition and C/O ratio (e.g. \citealt{2010ApJ...723.1195R}). Usually, C-rich Miras clearly change their mean magnitude over time, while the O-rich Miras usually do not show long-term irregular light variations. This is caused by the mass-loss phenomenon and the strong influence of circumstellar dust on stellar light \citep{2021ApJS..257...23I, 2022arXiv220300896O}.

A relatively short stage in stars life, the Mira-phase, is characterized by large bolometric luminosities \citep[ranging from several hundred to tens of thousands of the solar luminosity;][]{2021ApJS..257...23I} and low average densities \citep[about 7-8 orders of magnitude lower than the solar density;][]{2003A&A...397..943B, 2015pust.book.....C}. High-amplitude pulsations cause mass-loss phenomena via the stellar wind \citep{2020A&A...642A..82P}, which leads to an enrichment of the interstellar medium with elements heavier than nickel produced by the slow neutron-capture process \mbox{(s-process)}. Therefore Miras, like other LPVs, are important tracers of galactic chemistry \citep{1997MNRAS.288..512W, 2016A&A...586A..49B, 2019ApJ...887...82K, 2021MNRAS.501.5135Y}. 

Due to the large brightness and well-defined PLRs in NIR and mid-infrared wavelengths \citep[mid-IR;][]{2011MNRAS.412.2345I, 2018AJ....156..112Y, 2019ApJ...884...20B, 2020A&A...636A..48G}, Mira-type stars are promising distance indicators, and a suitable tool for studying stellar populations and structure of galaxies \citep{2009AcA....59..239S, 2011AcA....61..217S, 2013MNRAS.428.2216W, 2014EAS....67..263W, 2015MNRAS.452..910M, 2017AJ....153..170Y, 2018MNRAS.473..173W, 2018ApJ...857...67H, 2019MNRAS.483.5150M, 2020ApJ...889....5H, 2020MNRAS.492.3128G, 2020ApJ...891...50U}. 

Recently, \citet{2021ApJ...919...99I} calibrated PLRs in seven mid-IR bands from Wide-field Infrared Survey Explorer (WISE) and Spitzer Space Telescopes (wavelengths from $3.4$ $\mu$m to $22$ $\mu$m), separately for the O-rich and C-rich Miras. These PLRs allow measuring a distance to a single Mira with a precision at the level of 5\%. On the other hand, \citet{2021ApJS..257...23I} provided synthetic PLRs in 42 bands of existing and future sky surveys, that include e.g. The James Webb Space Telescope (JWST), The Nancy Grace Roman Space Telescope (formerly WFIRST), Vera C. Rubin Observatory (formerly LSST), the Hubble Space Telescope (HST) or The VISTA Near-Infrared $YJK_s$ Survey of the Magellanic Cloud System (VMC).

The growing number of known Miras, and therefore the ability to use them to study the structure of galaxies and population properties, is possible thanks to large-scale surveys providing multi-band time-series photometry or spectroscopy of variable stars. Among various sky surveys, one can distinguish, e.g., the Asteroid Terrestrial-impact Last Alert System (ATLAS, \citealt{2018AJ....156..241H}), the All Sky Automated Survey (ASAS; \citealt{1997AcA....47..467P}), the All-Sky Automated Survey for Supernovae (ASAS-SN; \citealt{2014ApJ...788...48S, 2019MNRAS.486.1907J}), the Catalina Sky Survey \citep[CSS;][]{2014ApJS..213....9D, 2017MNRAS.469.3688D}, Gaia \citep{2016A&A...595A...1G, 2018A&A...616A...1G, 2018A&A...618A..58M}, the Kilodegree Extremely Little Telescope (KELT; \citealt{2007PASP..119..923P, 2020ApJS..247...44A}), the Large Sky Area Multi-Object Fibre Spectroscopic Telescope (LAMOST; \citealt{2017ApJS..232...16Y}), the MAssive Compact Halo Objects (MACHO; \citealt{1995AJ....109.1653A, 2013OEJV..159....1B}), the  Panoramic Survey Telescope and Rapid Response System (Pan-STARRS; \citealt{2016arXiv161205560C, 2016arXiv161205243F}), the VISTA Variables in the Via Lactea (VVV; \citealt{2010NewA...15..433M}), the Zwicky Transient Facility (ZTF; \citealt{2019PASP..131a8002B, 2020ApJS..249...18C}), or finally the Optical Gravitational Lensing Experiment (OGLE; \citealt{2013AcA....63...21S, 2015AcA....65....1U}).

This paper focuses on almost 66,000 Mira-type stars found in the OGLE databases in the Galactic bulge and disk fields. In the International Variable Star Index  \citep[VSX;][]{2006SASS...25...47W}\footnote{https://www.aavso.org/vsx/, as of the day 30 March 2022} there are over 28,000 variable stars classified as Mira from the entire sky (type M). The vast majority of objects presented in this paper are new discoveries, which was possible thanks to the two-decade-long well-sampled high-quality time-series OGLE photometry. With this catalog, we triple the number of known Miras in the Milky Way. This collection of Miras will be a part of a much larger collection of LPVs found in the Galaxy, which will be a supplement for the LPVs collections found toward the Galactic bulge in the OGLE-III database \citep{2013AcA....63...21S}.

The paper is organized as follows. In Section \ref{section:data}, we describe the photometric data used in the analysis. In Section \ref{subsection:selection}, we discuss our methods of the selection and classification of Mira stars, but also we describe the method of period search, measurement of the mean magnitudes, and variability amplitude. Section \ref{section:catalog} presents the structure of the collection. In Section \ref{section:discussion}, we discuss the statistical properties of Mira-type variable stars in the Galactic bulge and disk. The paper is summarized in \mbox{Section \ref{section:conclusions}}. The long-term two-band time-series photometry together with finding charts for all stars are available from the OGLE Internet Archive\footnote{https://www.astrouw.edu.pl/ogle/}.

\begin{figure*}
\centering
\includegraphics[scale=0.25]{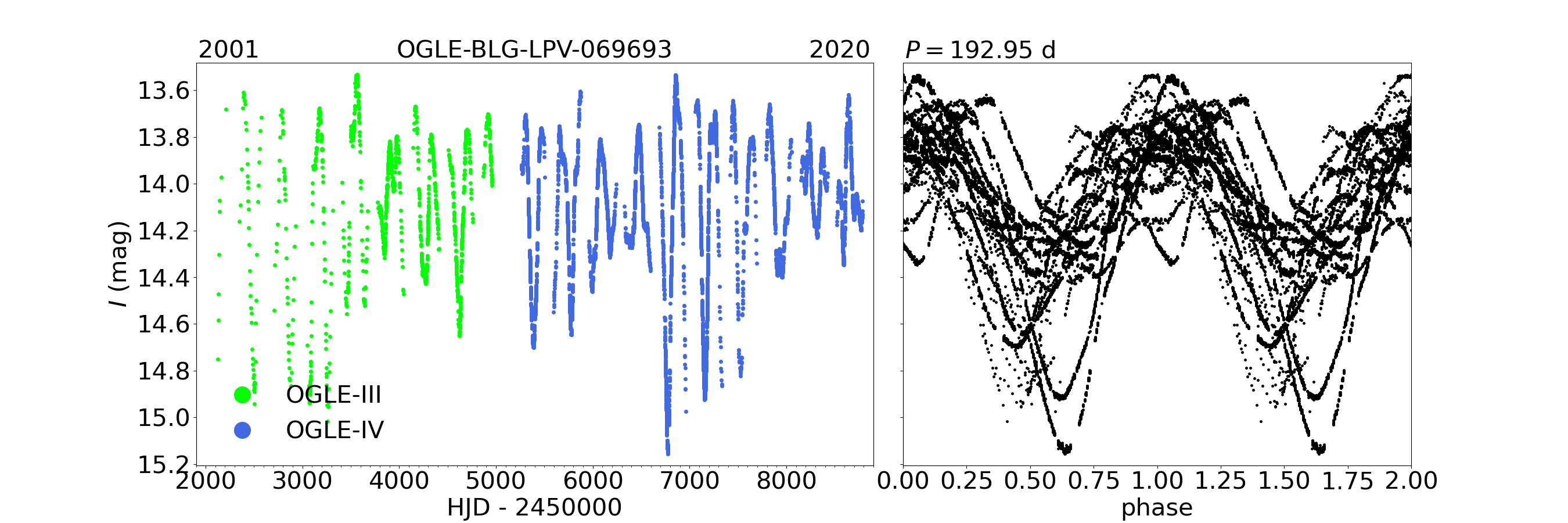}\\
\includegraphics[scale=0.25]{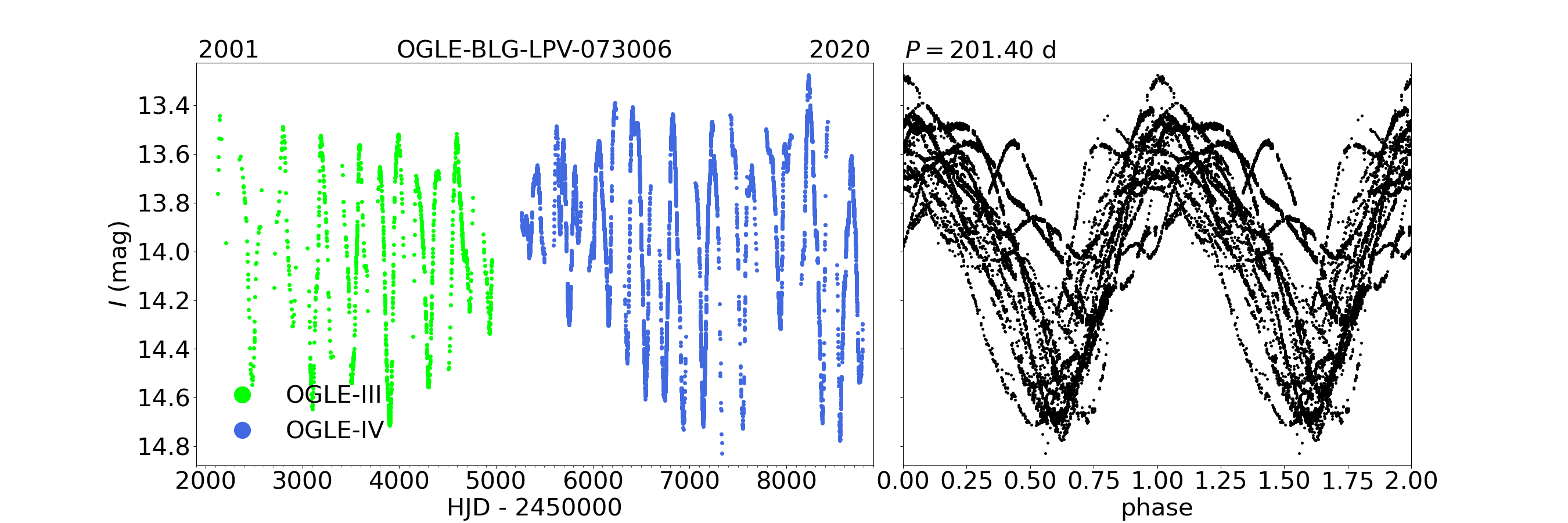}
\caption{Two examples of stars that were classified as Miras by \citet{2013AcA....63...21S}, but their amplitudes change significantly over time and fluctuate around $0.8$ mag. Such stars were reclassified from Mira-type to SRVs and excluded from the sample. Left-hand-side panels show unfolded light curves, while the right-hand-side panels show phase-folded light curves with pulsation periods $P$ (provided above the plots). The dates at the top of the unfolded light curves mark the year when the observations started and the year of the last used observations. The individual phases of the OGLE project are marked with different colors.}
\label{fig:mira_srv_reclassification}
\end{figure*}

\section{Observations and photometric data} \label{section:data}

The OGLE project is one of the longest-lasting variability sky survey worldwide. Historically, the first OGLE observations were taken in 1992 using the \mbox{1.0 m} Swope Telescope, located at the Las Campanas Observatory, Chile \citep[OGLE-I phase;][]{1992AcA....42..253U}. The observatory is operated by the Carnegie Institution for Science. The second phase of the OGLE survey (OGLE-II) started in 1997 at the Las Campanas Observatory with the new 1.3 m Warsaw Telescope dedicated to the project \citep{1997AcA....47..319U}. Since then, the regular monitoring of the Galactic bulge (BLG), Galactic disk (GD), and Magellanic Clouds has been carried out. During two decades, until 2020, the OGLE project went through two more phases: OGLE-III \citep[2001-2009;][]{2003AcA....53..291U}, and OGLE-IV \citep[2010-2020;][]{2015AcA....65....1U}. Between the phases, the CCD camera was upgraded, filling the entire 1.4 square degree field of view of the Warsaw Telescope in the OGLE-IV phase, with the pixel scale equal to $0.26''$. The two-decade-long monitoring of the southern sky was suspended in March 2020 due to the COVID-19 pandemic. Therefore, the last epoch in the collection was taken during the night of 17/18 March 2020.

The OGLE observations were carried out in the Johnson {\mbox{{\it V}-band}} (mean wavelength of $0.55$ $\mu$m) and Cousins {\it I}-band (mean wavelength of $0.81$ $\mu$m) filters. The vast majority of the photometric data were obtained in the {\it I}-band. Currently, less than half of all stars observed by the OGLE survey have any {\it V}-band observations. The exposure time in the {\it I}-band for the GD and outer BLG fields (outer parts of the BLG) was 25 seconds, while the typical exposure time in BLG fields was 100 seconds (mostly the inner part of the BLG). The magnitude range in the GD and outer BLG fields is a little wider than in the standard BLG survey, and spans from 10 mag and 20 mag, in contrast to the BLG fields with an exposure time of 100 seconds, where the magnitudes span from 13.5 mag to 21.5 mag. Observations in the \mbox{{\it V}-band} were taken less frequently, most often for the purposes of determining the $(V-I)$ color indices of stars. Therefore, the number of epochs in the \mbox{{\it V}-band} is on average an order of magnitude smaller than in the {\it I}-band. The exposure time in the {\it V}-band for GD and outer BLG fields was 25 seconds, while for inner BLG observations it was 150 seconds. For the {\it I}-band magnitude brighter than 15 mag, the uncertainty of a single photometric measurement is smaller than 0.01 mag, with a typical value of 0.005 mag, while stars with magnitudes ranging between 15 mag and 18 mag have photometric uncertainties smaller than 0.05 mag. In the \mbox{{\it V}-band}, stars brighter than 18 mag have photometric uncertainties smaller than 0.01 mag, with a typical value of \mbox{0.005 mag}. In total, the OGLE-IV survey covers $\sim$3000 square degrees of the sky. The OGLE-IV footprint of the Milky Way is presented in Figure \ref{fig:miras_map}.

The OGLE project uses the difference image analysis \citep[DIA;][]{1998ApJ...503..325A, 2000AcA....50..421W} to obtain photometry of the sources. For more information about technical details, instrumentation, data processing, sky coverage, or calibration in each OGLE phase, we encourage the reader to see the papers by \citet{1992AcA....42..253U, 1997AcA....47..319U, 2003AcA....53..291U, 2008AcA....58...69U, 2015AcA....65....1U}.

\newpage
\section{Selection and classification of Miras} \label{subsection:selection}

To date, only one catalog of LPVs toward the BLG has been published based on the OGLE-III data \citep{2013AcA....63...21S}. Among $\sim$230,000 LPVs, the authors found 6528 Mira-type variables. The vast majority of the sample consists of the OGLE small amplitude red giants (OSARGs; over 190,000 objects). A search for LPVs in the GD fields has never been carried out based on the OGLE data, however, 31 LPVs were found during the analysis of a OGLE-III GD area of 7.12 square degrees in the direction tangent to the Centaurus Arm \citep{2013AcA....63..379P}.

In this paper, we are interested in finding Mira-type variable stars. Miras variability is characterized by a large brightness amplitude \citep[conventionally equal to or greater than 0.8 mag in the {\it I}-band;][]{2013AcA....63...21S} and long pulsation periods (from about $80$ days to more than $1000$ days). Moreover, Miras pulsate solely in the fundamental mode, so a single significant period is expected.

The search for Miras began with selecting over a billion point sources in the BLG and GD fields, from which we excluded already classified and published Miras \citep{2013AcA....63...21S, 2013AcA....63..379P}. In the second step, we subtracted long-term trends estimated with the cubic splines from the {\it I}-band OGLE-IV light curves to properly determine the amplitude of brightness variations. After that, we calculated the \mbox{{\it I}-band} brightness amplitude defined as the difference between the faintest and brightest data points, rejecting $2.5\%$ points from both sides of the brightness distribution to avoid amplitude contamination by outlying measurements. Then, we measured the change of dynamics in the brightness between consecutive, closest measurements given by the equation:

\begin{equation}
     c = \frac{A_I}{\frac{1}{n}\sum_{i=1}^{n-1}\frac{\left|I(t_{i+1})-I(t_i)\right|}{t_{i+1}-t_i}},
\label{eqn:dynamics}
\end{equation}

\noindent where $A_I$ is the {\it I}-band brightness amplitude, $I(t_{i+1})$ and $I(t_i)$ are the {\it I}-band brightness of two consecutive measurements, and $t_{i+1} - t_i$ is the time difference between these measurements. This coefficient was calculated using all $n$ points of light curves taking into account only these pairs of measurements, for which the time difference was smaller than 10 days (i.e., \mbox{$(t_{i+1} - t_i) < 10$} d). The larger the coefficient $c$, the greater the probability that the analyzed star is of Mira-type (or LPV in general). We randomly selected several dozen light curves from many different ranges of this coefficient and visually inspected each of them. We noticed that if the coefficient $c$ is below 15 days, we no longer detected Miras in the sample. Out of all light curves, we accepted those with a coefficient $c$ greater than or equal to 15 days only. This limit was chosen experimentally and it is arbitrary. Moreover, we made an amplitude cut, and we removed all objects with the brightness amplitude below 0.5 mag. This left us with 194,522 light curves for further analyses.

For the selected light curves we calculated periods using two methods: the Lomb-Scargle periodogram \citep[LS;][]{1976Ap&SS..39..447L, 1982ApJ...263..835S, 2018ApJS..236...16V}, and the {\sc FNPEAKS} code\footnote{\url{http://helas.astro.uni.wroc.pl/deliverables.php?active=fnpeaks}} (created by Z. Ko\l{}aczkowski, \mbox{W. Hebisch}, and G. Kopacki, 2003), based on the standard discrete Fourier transform modified for unevenly spaced data \citep{1985MNRAS.213..773K}. A detailed discussion on period searching methods for such data can be found in \citet{2019ApJ...879..114I}. 

We searched the frequency space from $0.0005$ to \mbox{$0.1$ day$^{-1}$} (periods from 10 to 2000 days), with a resolution of $10^{-6}$ day$^{-1}$. At this stage of the analysis, the most arduous and lengthy process began, i.e., the visual inspection and classification of stars. The selection and classification processes based on the visual inspection of light curves by experienced astronomers guarantee the high purity and completeness of our catalogs. Our priority is to find all objects of a given type in the OGLE databases, at the same time, making sure not to contaminate the sample with other types of variability.

We checked by eye both raw (unfolded) and folded light curves. The light curves of stars classified as Miras were checked after being folded with each of the measured periods (from the {\sc FNPEAKS} and LS method). If both periods were similar, we chose the one that phased the light curve better, i.e., gave a smaller scatter of points. If one of the periods was correct, we left it without any changes. In the case when both periods were incorrect, we corrected the period manually. In this step, we also removed obvious outlying measurements.
Here we classified 60,997 stars as Mira candidates. More than half of the examined light curves were classified as artifacts, e.g., caused by bright, saturated stars. We also found a few thousand SRVs, a few thousand stars with other variability types, a few hundred long secondary period variables \citep[see, e.g.,][]{2021ApJ...911L..22S}, and finally several microlensing events.

Some OGLE fields partially overlap, so stars located on the edges of the fields are measured twice, and have more than one light curve in our database. Such pairs of stars could be used to estimate the completeness of the catalog (for a detailed description see Section \ref{subsection:completeness}). We found 2630 pairs of stars in the overlapping parts of adjacent OGLE fields. Typically, stars in these pairs differ in the number of measurements and the quality of measurements (i.e., measurements uncertainties). We visually inspected each pair and we chose that element of the pair that had a better light curve in the sense of measurements uncertainty and the number of data points. After that, the number of Mira candidates dropped to 58,367.

\subsection{Determination of final periods, mean magnitudes, and amplitudes} \label{subsection:final_measurements}

Periods measured for stars classified as Miras, although manually corrected and verified by eye, may still be inaccurate. We decided to fine-tune periods by using the {\sc TATRY} code, which implements the multi-harmonic analysis of variance algorithm \citep[ANOVA;][]{1996ApJ...460L.107S}. We did not use the {\sc TATRY} code for the initial period determination because this method requires more computational time comparing to the used ones. For most stars, the period found in this way was very close to that accepted during the visual inspection of the light curves. In such cases, we chose this period as the final one. However, in some cases, the differences between periods were significant. Therefore, we verified the light curves phase-folded with both periods and applied the period that phased the light curve best. For period fine-tuning we used all available OGLE data.

After the pulsation periods fine-tuning, we decided to fine-tune the brightness variation amplitude. We phase-folded each light curve with the best period. As OGLE light curves may be unevenly sampled, we filled the gaps between the data points using the cubic spline interpolation. Then, we fitted the third-order truncated Fourier series using the measured period to the phase-folded ``filled'' light curve, and we measured the peak-to-peak amplitude. A similar procedure was applied to the mean magnitude measurements, with the difference that we fitted the third-order Fourier series to a light curve transformed to the flux scale. We integrated the fit to determine the mean brightness and finally, we transformed the mean brightness to the magnitude scale. The {\it I}-band amplitude and both {\it I}- and {\it V}-band mean magnitudes were measured based on the {\mbox{OGLE-IV}} data. In the case that data from the fourth phase of the OGLE project were not available, the measurements were made using data from the previous phases of the survey.

\subsection{Amplitude cut} \label{subsection:amplitude_cut}

\citet{2013AcA....63...21S} adopted the {\it I}-band amplitude division line between SRVs and Miras equal to 0.8 mag (see Figure 8 therein). We adopted the same division value to remove stars with {\it I}-band amplitude smaller than 0.8~mag. This step left us with 58,047 Mira-type variable stars. The removed stars were classified as SRVs and will be published in the full version of the OGLE collection of LPVs.

This amplitude limit is conventional, and stars with amplitudes close to $0.8$ mag may be misclassified. 
On the other hand, we found stars that change amplitudes and their classifications change over time from Miras to SRVs and vice versa. Such cases are described in Section \ref{subsection:reexamined}.

\begin{table*}[ht!]
\caption{Table \texttt{ident.dat} with identifications and equatorial coordinates for all Miras.}
\tiny
\begin{center}
\begin{tabular}{lcccccccc}
\hline \hline
ID & Type & Loc. & R.A. & Decl. & OGLE-IV ID & OGLE-III ID & OGLE-II ID & Other ID \\
 &   &  &  (h:m:s) & (deg:m:s) &  &  &  &  \\ \hline
OGLE-BLG-LPV-000009 & Mira & BLG & 17:05:28.47 & $-$32:44:22.4 & BLG897.03.15 & BLG366.6.6796 & -- & Terz V 3396 \\
OGLE-BLG-LPV-000018 & Mira & BLG & 17:05:37.00 & $-$33:02:36.6 & BLG898.27.195 & BLG366.8.10859 & -- & -- \\
OGLE-BLG-LPV-000024 & Mira & BLG & 17:05:43.89 & $-$33:02:32.6 & BLG898.27.79 & BLG366.8.10875 & -- & --\\
OGLE-BLG-LPV-000028 & Mira & BLG & 17:05:52.29 & $-$32:39:49.4 & BLG897.02.55701 & BLG366.6.69850 & -- & --\\
OGLE-BLG-LPV-000030 & Mira & BLG & 17:05:55.21 & $-$32:50:24.6 & BLG898.27.60162 & BLG366.7.42260 & -- & Terz V 3407\\
OGLE-BLG-LPV-000092 & Mira & BLG & 17:07:36.82 & $-$32:48:40.3 & BLG908.16.54571 & BLG366.2.82751 & -- & Terz V 3484\\
OGLE-BLG-LPV-000106 & Mira & BLG & 17:07:44.70 & $-$32:30:20.8 & BLG908.24.58094 & BLG366.4.89376 & -- & Terz V 3493\\
OGLE-BLG-LPV-000113 & Mira & BLG & 17:07:50.52 & $-$32:43:04.1 & BLG908.24.73 & BLG366.3.26435 & -- & Terz V 3498\\
OGLE-BLG-LPV-000142 & Mira & BLG & 17:08:39.85 & $-$32:58:09.9 & BLG908.14.267 & BLG332.6.12 & -- & --\\
OGLE-BLG-LPV-000147 & Mira & BLG & 17:08:40.98 & $-$32:54:57.2 & BLG908.14.58864 & BLG332.6.61774 & -- & Terz V 3538\\
OGLE-GD-LPV-025636 & Mira & GD & 19:16:07.04 & $-$00:06:14.3 & DG1071.17.12320 & -- & -- & ASASSN-V J191607.04-000614.3 \\ \hline
\end{tabular}
\end{center}
\label{table:ident.dat}
\tablecomments{For each star, we provide its unique ID, variability type (the same for each object; the Miras collection will be part of a much larger collection of LPVs in the Milky Way), J2000 equatorial coordinates, OGLE-IV, OGLE-III, and OGLE-II identifiers, and also identifiers from other catalogs (the best match), i.e. VSX/ASAS-SN/ZTF catalogs. The '--' sign means a lack of match with other databases. \\ (This table is available in its entirety in a machine-readable form.)}
\end{table*}

\begin{table*}[ht!]
\caption{Table \texttt{Miras.dat} with observational parameters of each star: the {\it I}- and {\it V}-band mean magnitudes, pulsation period, and the {\it I}-band amplitude.}
\footnotesize
\begin{center}
\begin{tabular}{lccccc}
\hline \hline
ID & Loc. & {\it I} & {\it V} & $P$ & $A_I$ \\
 &   & (mag) & (mag) & (days)  & (mag)  \\ \hline
OGLE-BLG-LPV-000009 & BLG & 13.008 & - & 275.3 & 2.345 \\
OGLE-BLG-LPV-000018 & BLG & 13.568 & - & 363.9 & 2.333 \\
OGLE-BLG-LPV-000024 & BLG & 12.718 & - & 245.7 & 1.848 \\
OGLE-BLG-LPV-000028 & BLG & 12.868 & - & 386.8 & 1.488 \\
OGLE-BLG-LPV-000030 & BLG & 12.133 & - & 172.01 & 1.926  \\
OGLE-BLG-LPV-000092 & BLG & 12.980 & - & 348.1 & 2.145 \\
OGLE-BLG-LPV-000106 & BLG & 13.051 & - & 363.1 & 2.764 \\
OGLE-BLG-LPV-000113 & BLG & 13.315 & - & 342.1 & 1.428 \\
OGLE-BLG-LPV-000142 & BLG & 13.190 & - & 268.1 & 2.626 \\
OGLE-BLG-LPV-000147 & BLG & 12.493 & - & 208.1 & 1.774 \\
OGLE-GD-LPV-025636 & GD & 12.207 & - & 255.5 & 3.257 \\ \hline
\end{tabular}
\end{center}
\label{table:miras.dat}
\tablecomments{For each star, we provide the {\it I}-band and {\it V}-band (if available; in total {\it V}-band light curves are available for 19,779 stars) mean magnitudes, pulsation period, and the {\it I}-band amplitude. The '--' sign for the {\it V}-band mean magnitude means that the {\it V}-band light curve is not available. \\ (This table is available in its entirety in a machine-readable form.)}
\end{table*}

\subsection{Re-examining of the OGLE-III catalog} \label{subsection:reexamined}

In the OGLE-III catalog of LPVs toward the BLG, \citet{2013AcA....63...21S} classified 6528 stars as Mira-type. We re-examined all these stars based on the OGLE-IV light curves. We applied methods described in Sections \ref{subsection:final_measurements} and \ref{subsection:amplitude_cut}, and we visually inspected all these light curves. We found that the amplitudes of 112 stars were changing over time and were fluctuating around 0.8 mag. For such stars, we decided to change the classification from Mira to SRVs, and we excluded them from the sample. In Figure \ref{fig:mira_srv_reclassification}, we present two examples of stars that change the brightness variation amplitude significantly over time, thus the classification of such stars is not obvious. Stars 'switching' between the Mira-type and SRVs are important from the evolutionary point of view because SRVs are believed to be the progenitors of Miras \citep[e.g.][]{2022A&A...658L...1T}, so they should be examined in detail.

Consequently, we were left with 6416 Miras from the \mbox{OGLE-III} catalog \citep{2013AcA....63...21S} and we supplemented our sample with these stars. The number of Miras increased to 64,463.

\subsection{Cross-matching with other catalogs of variable stars} \label{subsection:crossmatching}

We crossmatched our sample with the VSX \citep[][]{2006SASS...25...47W}\footnote{https://www.aavso.org/vsx/index.php}, the ASAS-SN variable stars catalog \citep{2014ApJ...788...48S, 2019MNRAS.486.1907J}, and the ZTF variable stars catalog \citep{2019PASP..131a8002B, 2020ApJS..249...18C}. We checked how many stars from these sources are in the OGLE fields, but were not found during our search. We identified 3275 such objects. We searched the OGLE databases for objects in a radius of 1 arcsec around the coordinates from the above-mentioned catalogs. We extracted all available light curves, measured the pulsation periods and amplitudes for these stars (as described in Section \ref{subsection:final_measurements}) and we carefully checked each of them.
As a result, we supplemented our catalog with 1518 missed Miras, finally closing our search for Mira-type variables in the Milky Way. The remaining 1757 stars turned out to be variables of other types (mainly SRVs), were too bright to be observed by the OGLE survey, or we did not find a counterpart in the OGLE database, probably due to inaccurate coordinates provided in the external catalogs. In total, our collection consists of 65,981 Mira variables. The distribution of Miras in the sky is presented in Figure \ref{fig:miras_map}.

\begin{figure*}
\centering
\includegraphics[scale=0.2]{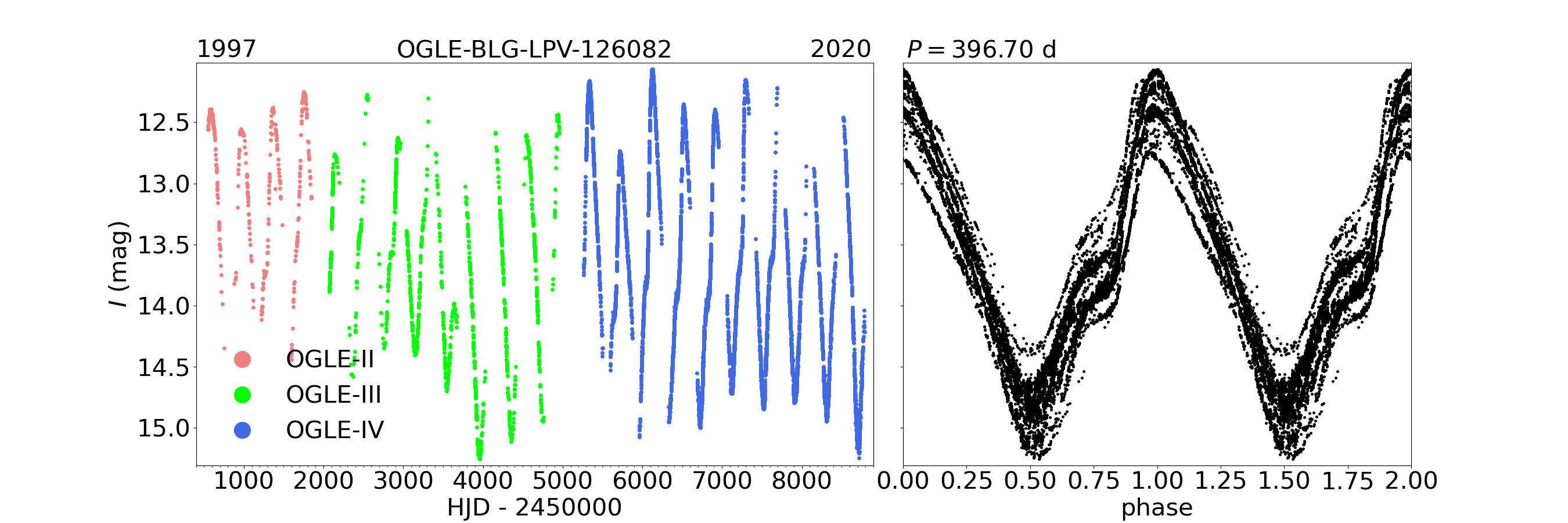}
\includegraphics[scale=0.5]{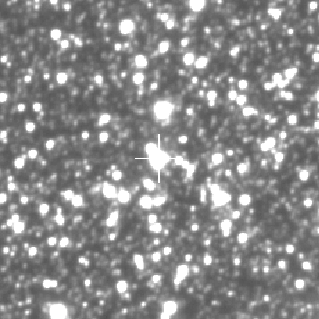}\\
\includegraphics[scale=0.2]{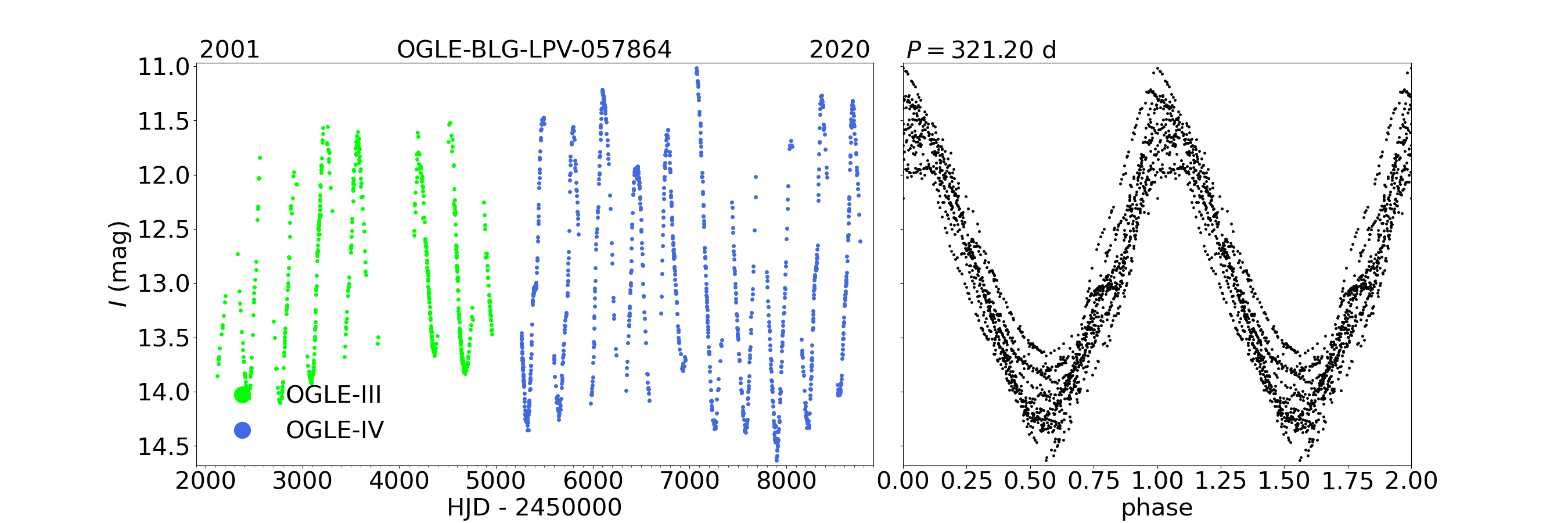}
\includegraphics[scale=0.5]{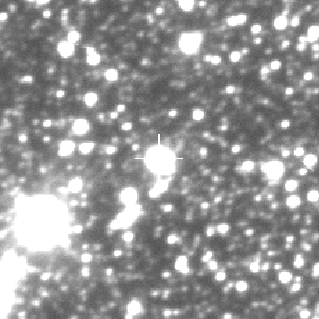}\\
\includegraphics[scale=0.2]{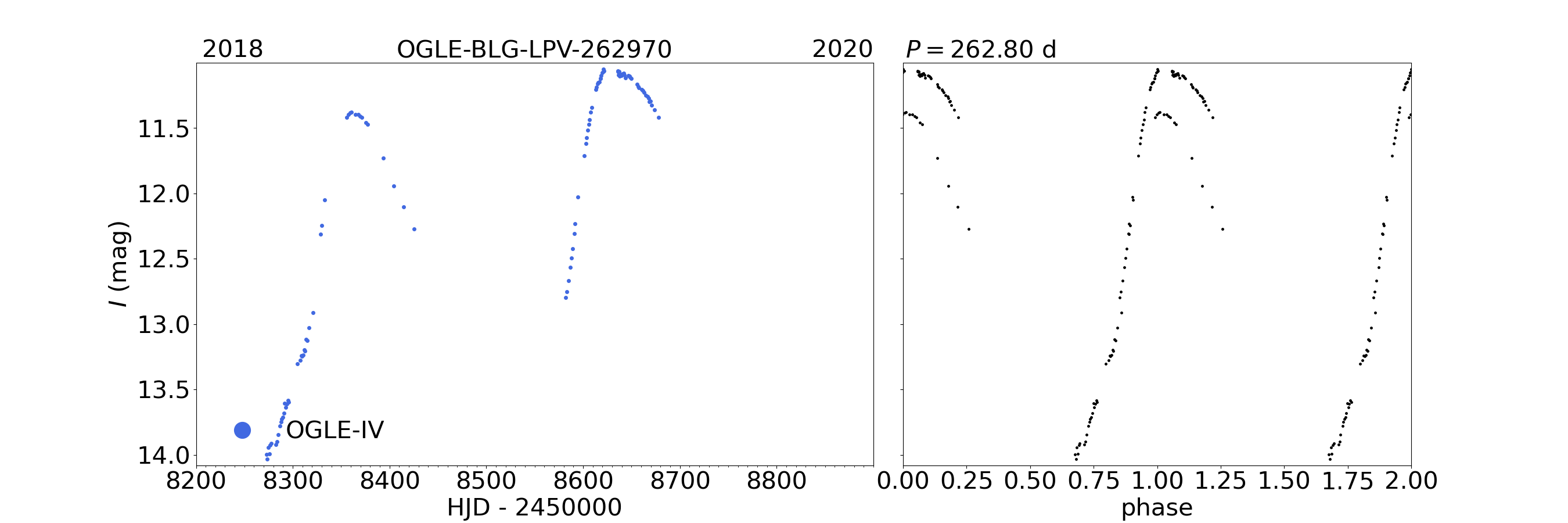}
\includegraphics[scale=0.5]{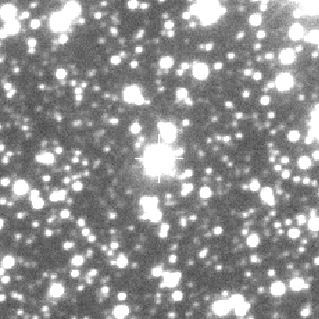}\\
\caption{Three examples of Miras from our collection, observed in the BLG fields, with light curves containing a different number of epochs. {\it Top row:} A well-covered light curve of Mira OGLE-BLG-LPV-126082. The observations consist of 19,169 epochs that span 23 years (from 1997 to 2020). {\it Middle row:} A medium-covered light curve of Mira OGLE-BLG-LPV-057864. The observations consist of 1593 epochs that span 19 years (from 2001 to 2020). {\it Bottom row:} A poorly-covered light curve of Mira OGLE-BLG-LPV-262970. The observations consist of 119 epochs that span 2 years (from 2018 to 2020). Left-hand-side panels show the unfolded light curves, while the right-hand-side panels show phase-folded light curves with pulsation periods $P$ (provided above the plots). The dates at the top of the unfolded light curve mark the year when the observations started and the year of the last used observations. The phases of the OGLE project are marked with different colors (red, green, and blue). For each star, we also provide the $60'' \times 60''$ finding chart oriented with the North up and East to the left. Each Mira is at the center of the finding chart and is marked with a white cross.}
\label{fig:BLG_lc}
\end{figure*}

\begin{figure*}
\centering
\includegraphics[scale=0.2]{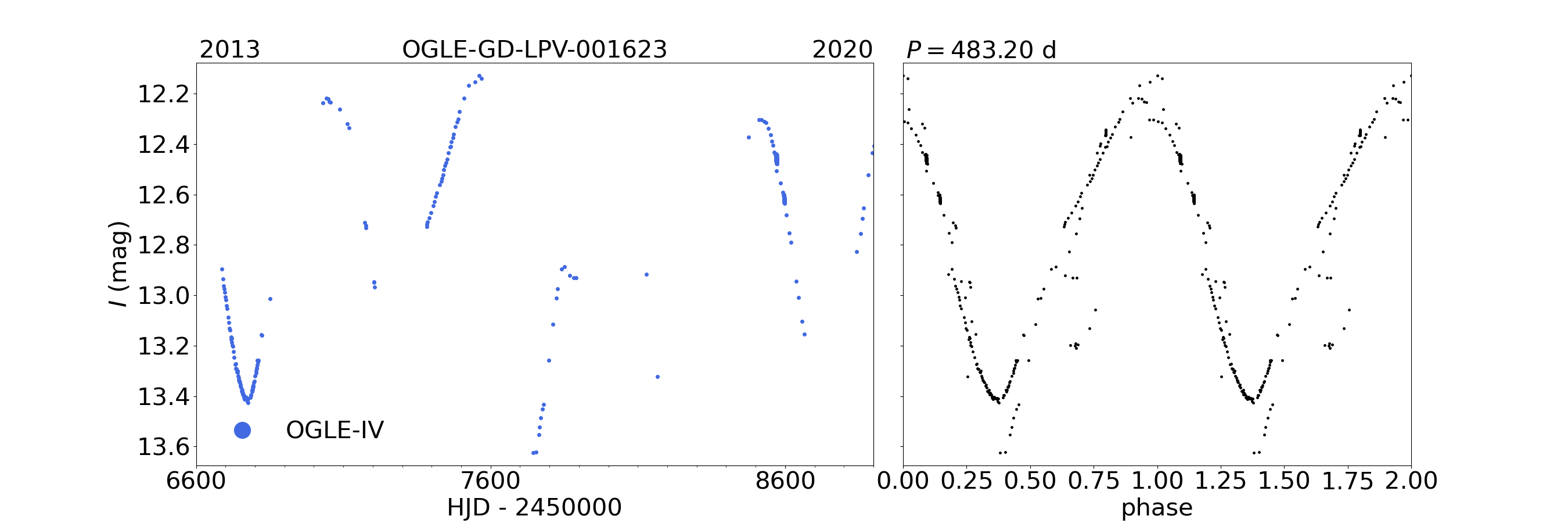}
\includegraphics[scale=0.5]{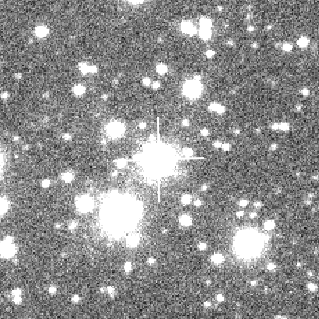}\\
\includegraphics[scale=0.2]{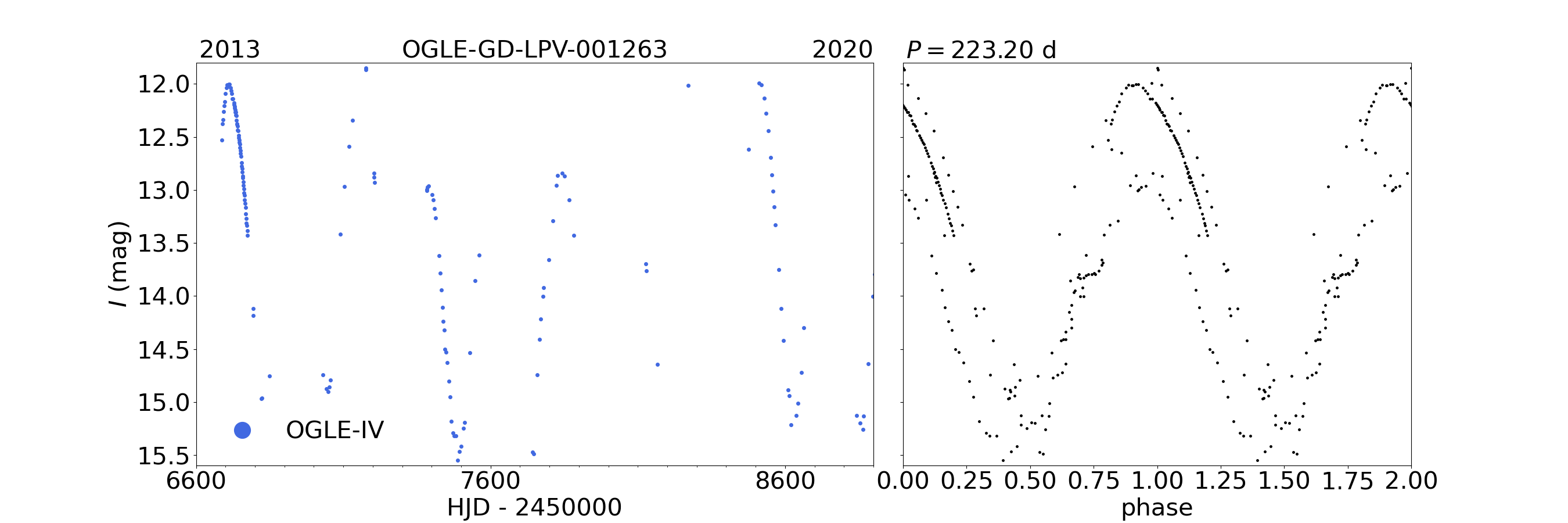}
\includegraphics[scale=0.5]{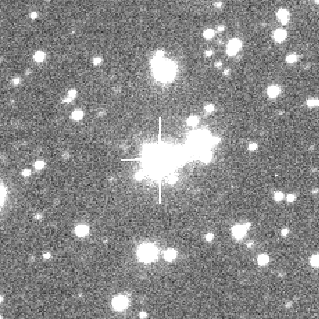}\\
\includegraphics[scale=0.2]{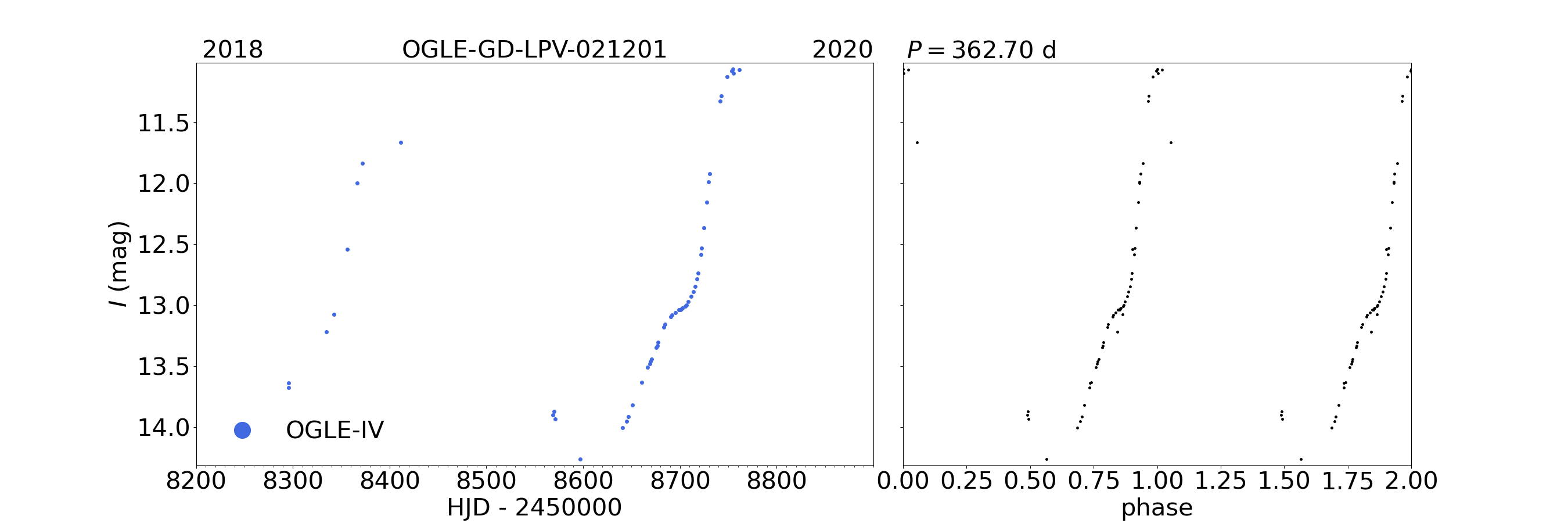}
\includegraphics[scale=0.5]{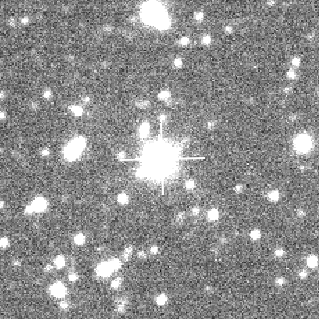}\\
\caption{Three examples of Miras from our collection, observed in the GD fields, with light curves containing a different number of epochs. {\it Top row:} A well-covered light curve of Mira OGLE-GD-LPV-001623. The observations consist of 1307 epochs that span 7 years (from 2013 to 2020). {\it Middle row:} A medium-covered light curve of Mira OGLE-GD-LPV-001263. The observations consist of 180 epochs that span 7 years (from 2013 to 2020). {\it Bottom row:} A poorly-covered light curve of Mira OGLE-GD-LPV-021201. The observations consist of 55 epochs than span 2 years (from 2018 to 2020). Left-hand-side panels show unfolded light curves, while the right-hand-side panels show phase-folded light curves with pulsation periods $P$ (provided above the plots). The dates at the top of unfolded light curve mark the year when the observations started and the year of the last used observations. The fourth phase of the OGLE project is marked with blue color. For each star, we also provide the $60'' \times 60''$ finding chart oriented with the North up and East to the left. Each Mira is at the center of the finding chart and is marked with a white cross.}
\label{fig:GD_lc}
\end{figure*}

\section{The OGLE Collection of Mira Variables} \label{section:catalog}

We present the OGLE collection of Mira stars in the Milky Way, which contains 65,981 objects in total. This is the largest, purest, and most complete catalog of Mira-type variables to date. The data for all these stars are available through the OGLE website and FTP:
\begin{itemize}
    \item \url{https://www.astrouw.edu.pl/ogle/ogle4/OCVS/blg/lpv/} -- objects in the BLG fields,
    \item \url{https://www.astrouw.edu.pl/ogle/ogle4/OCVS/gd/lpv/} -- objects in the GD fields,
    \item \url{ftp.astrouw.edu.pl/ogle/} -- FTP for the entire OGLE Collection of Variable Stars,
    \item \url{https://ogledb.astrouw.edu.pl/~ogle/OCVS/} -- user-friendly search engine for the entire OGLE Collection of Variable Stars.
\end{itemize}

Each object has an individual identifier: OGLE-BLG-LPV-NNNNNN or OGLE-GD-LPV-NNNNNN, where NNNNNN is a six-digit number. The identifiers of stars in the BLG fields remain unchanged with respect to the collection published by \citet{2013AcA....63...21S}. Stars discovered in the OGLE-IV data receive new identifiers and are listed in an ascending right ascension order. To date, only 31 LPVs (18 Miras and 13 SRVs) from the GD area were discovered in the OGLE data \citep{2013AcA....63..379P}. Our collection of Miras contains 20 objects from the \citet{2013AcA....63..379P} catalog (two stars previously categorized as SRVs have been reclassified as Miras). Their designations are the same as in the OGLE-III catalog, but we increased the number of digits in the identifiers from four to six. In the future, we plan to publish a much larger collection of LPVs in the Milky Way, of which this Mira stars collection will be part.

\begin{table}[]
\caption{The median (Med.) and maximum (Max.) number of epochs in the light curves, separately for each OGLE phase, and with division into locations and bands. We also provide fraction of Miras that have light curves in the indicated OGLE phases and bands, relative to the total number of Miras in the BLG (40,356 stars), and in the GD fields (25,625 star).}
\begin{center}
\begin{tabular}{cccccc}
\hline \hline
Phase & Band & Location & Med. & Max. & \% of Miras \\ \hline
OGLE-II &  &   & 310 & 550 & 4\% \\
OGLE-III & {\it I} & BLG  & 623 & 2532 & 15\% \\
OGLE-IV &  &   & 227 & 16,686 & 97\% \\ \hline
OGLE-II &  &   & 9 & 18 & 3\%\\
OGLE-III & {\it V} & BLG  & 5 & 41 & 10\% \\
OGLE-IV &  &  & 42 & 230 & 27\% \\ \hline
OGLE-II &  &  & 125 & 553 & 1\% \\
OGLE-III & {\it I} & GD  & 1676 & 2696 & 0.1\% \\
OGLE-IV &  &   & 130 & 1307 & 99\% \\ \hline
OGLE-II &  &   & 69 & 98 & 0.01\% \\
OGLE-III & {\it V} & GD  & 4 & 8 & 0.06\% \\
OGLE-IV &  &   & 8 & 16 & 30\% \\
\hline
\end{tabular}
\end{center}
\label{table:lc_statistics}
\end{table}

The websites of the OGLE collection of Miras are structured as follows. The file \texttt{ident.dat} contains the list of stars with classification, their J2000 equatorial coordinates, identifications in the OGLE-II, OGLE-III, and OGLE-IV databases, and designations taken from the VSX/ASAS-SN/ZTF catalogs (in total, we identified 12,033 Miras in external catalogs of variable stars). Identifier of Miras in the OGLE databases contains shortcuts of the OGLE fields identifying the position of stars in the sky: BLG is for stars located in the inner and outer Galactic bulge fields, GD is for stars located in the Galactic disk fields with a negative Galactic longitude, whereas DG stands for stars located in the Galactic disk fields, with the positive Galactic longitude. In the file \texttt{Miras.dat}, we provide observational parameters for each star -- the pulsation period, mean magnitudes in {\mbox{{{\it I}-} and {\it V}-bands}}, and {\it I}-band brightness amplitude. The files \texttt{ident.dat} and \texttt{Miras.dat} are also provided with this paper in a machine-readable form (merged for both environments). In Table \ref{table:ident.dat} and Table \ref{table:miras.dat}, we present the first ten rows, and the last row from tables \texttt{ident.dat} and \texttt{Miras.dat} for guidance regarding their forms and content.

The subdirectory \texttt{fcharts/} contains finding charts for all objects. We provide $60'' \times 60''$ subframes of the $I-$band DIA reference images, oriented with the North up, and East to the left. The object is marked with a white cross and is in the center of the frame.

In subdirectories \texttt{phot\_ogle2/}, \texttt{phot\_ogle3/}, \texttt{phot\_ogle4/}, we provide the {\it I}-band and {\it V}-band (if available) time-series photometry from the OGLE-II (1997-2000), OGLE-III (2001-2009), and OGLE-IV (2010-2020) phases, respectively. All available data from the OGLE photometric databases are included here. The data from different phases were calibrated separately to the standard Johnson-Cousins photometric system. During our classification, we removed obvious outlying points from the light curves. We encourage the reader to pay attention to possible offsets between 
photometric zero points for the light curves obtained during different 
phases of the OGLE project. The typical uncertainties of the OGLE 
photometric calibrations do not exceed 0.05~mag, however, in individual 
cases much larger offsets may occur, which can be a result of several 
factors, e.g., high local star density which causes blending. Therefore, significant offsets should be taken into account during the light curves merging process (if needed).

The {\it I}-band light curves (from at least one of the OGLE phases) are available for all Mira stars in our collection. In turn, the {\it V}-band observations are available only for 19,779 stars (11,883 from BLG, and 7896 from GD). The epoch number statistics of the OGLE light curves, separately for BLG and GD, and with division into {\mbox{{\it I}- and {\it V}-band}} are presented in Table \ref{table:lc_statistics}. In Figures \ref{fig:BLG_lc} and \ref{fig:GD_lc}, we present three examples of Miras observed in the BLG (Fig.~\ref{fig:BLG_lc}), and in the GD fields (Fig.~\ref{fig:GD_lc}). Each figure presents light curves from the well-covered (the largest number of measurements, top panel), to the poorly-covered one (bottom panel). In these figures, we also present finding charts for each star. 

\begin{figure*}
\centering
\includegraphics[scale=0.65]{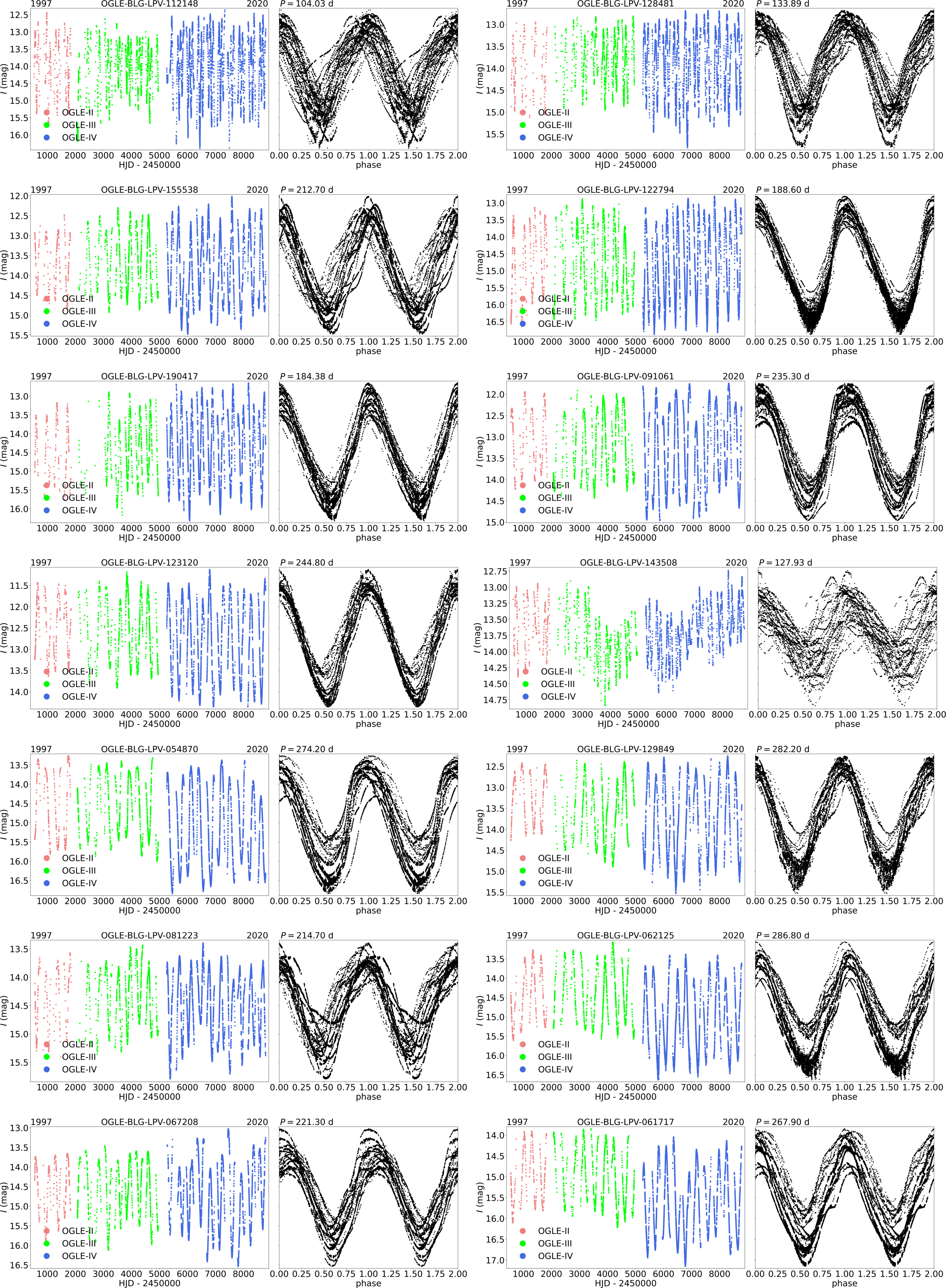}
\caption{Fourteen examples of Mira stars from our catalog which were observed since 1997, i.e., the beginning of the OGLE-II phase, until March 2020. Left-hand-side panels show unfolded light curves, while the right-hand-side panels show phase-folded light curves with pulsation periods $P$ (provided above the plots). The dates at the top of the unfolded light curve mark the year when the observations started and the year of the last used observations. The individual phases of the OGLE project are marked with different colors (red, green, and blue).}
\label{fig:big_lc_1}
\end{figure*}

\begin{figure*}
\centering
\includegraphics[scale=0.65]{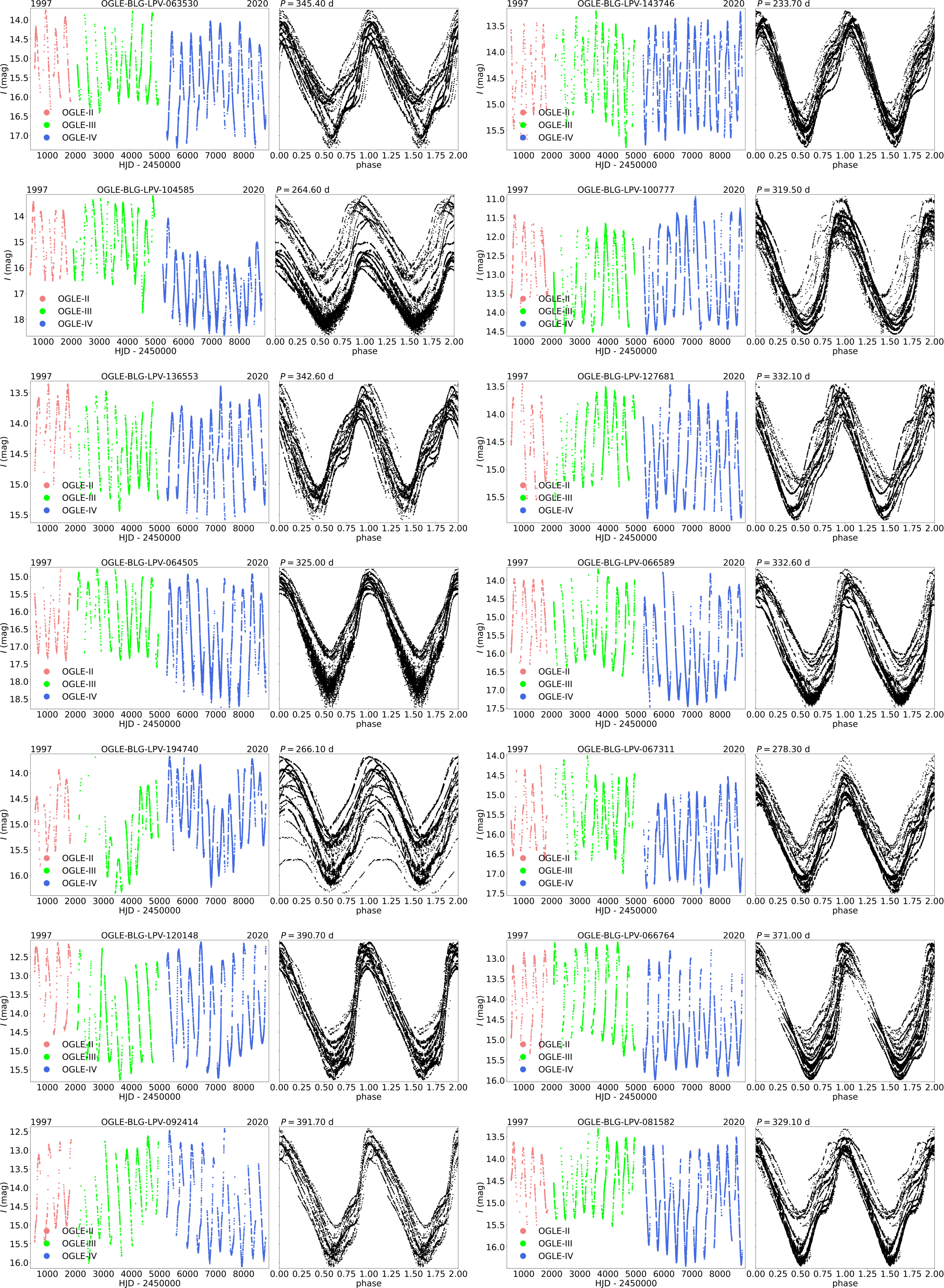}
\caption{Same as Figure \ref{fig:big_lc_1}, but fourteen other Miras are presented.}
\label{fig:big_lc_2}
\end{figure*}

\begin{figure*}
\centering
\includegraphics[scale=0.65]{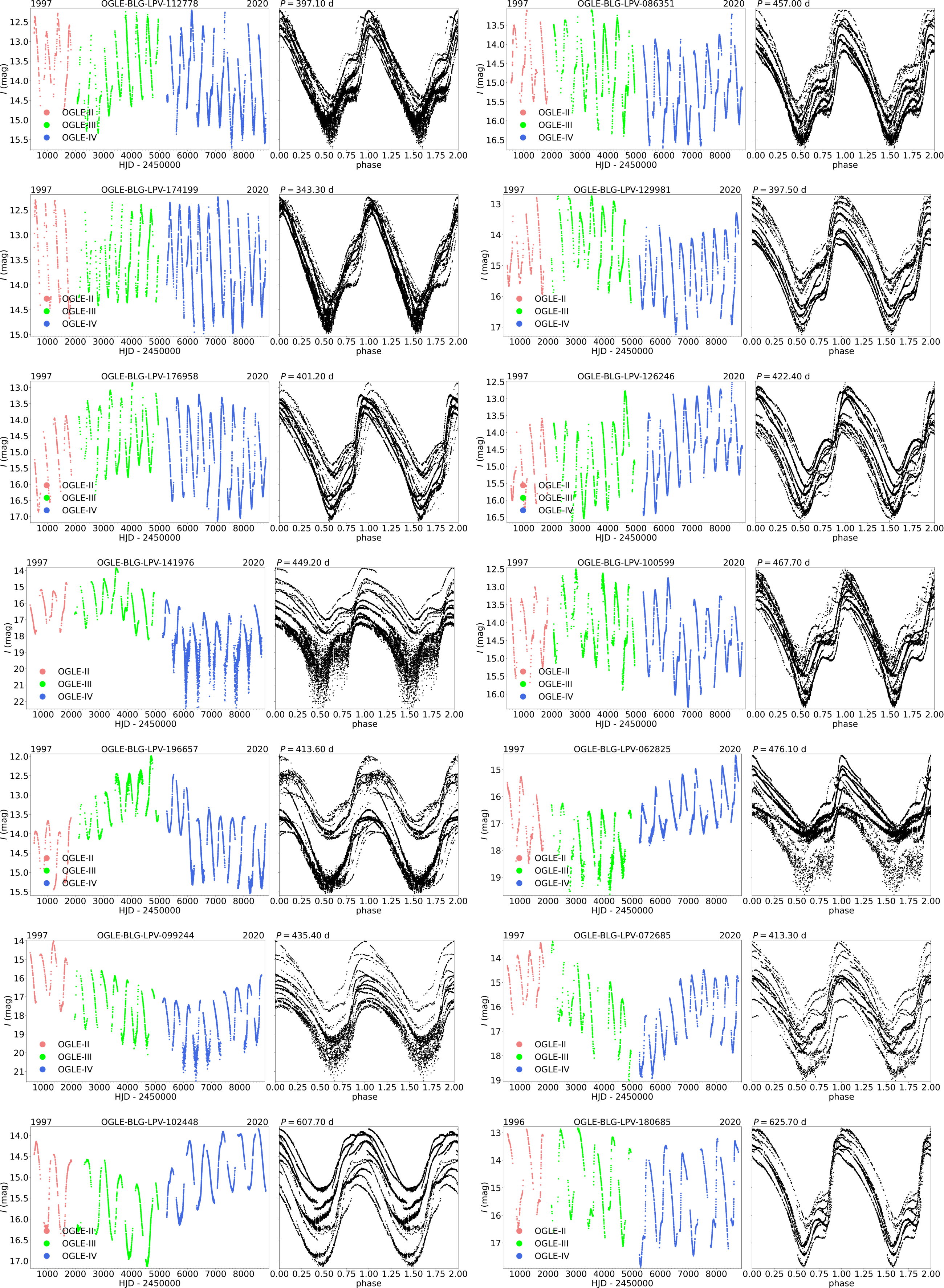}
\caption{Same as Figure \ref{fig:big_lc_1} and Figure \ref{fig:big_lc_2}, but fourteen other Miras are presented.}

\label{fig:big_lc_3}
\end{figure*}

Our collection includes 939 Miras observed continuously from 1997, i.e., from the beginning of the OGLE-II phase, until March 2020 when the telescope had to be closed due to the COVID-19 pandemic. This dataset contains objects that significantly change their mean magnitudes or brightness amplitude over time, or the pulsation period lengthens or shortens.
This is a unique dataset which can be used to study, e.g., the evolution of Miras or the mass-loss phenomenon. Examples of 42 light curves of such objects are presented in Figures \ref{fig:big_lc_1}, \ref{fig:big_lc_2}, and \ref{fig:big_lc_3}.

\begin{figure}
\centering
\includegraphics[scale=0.37]{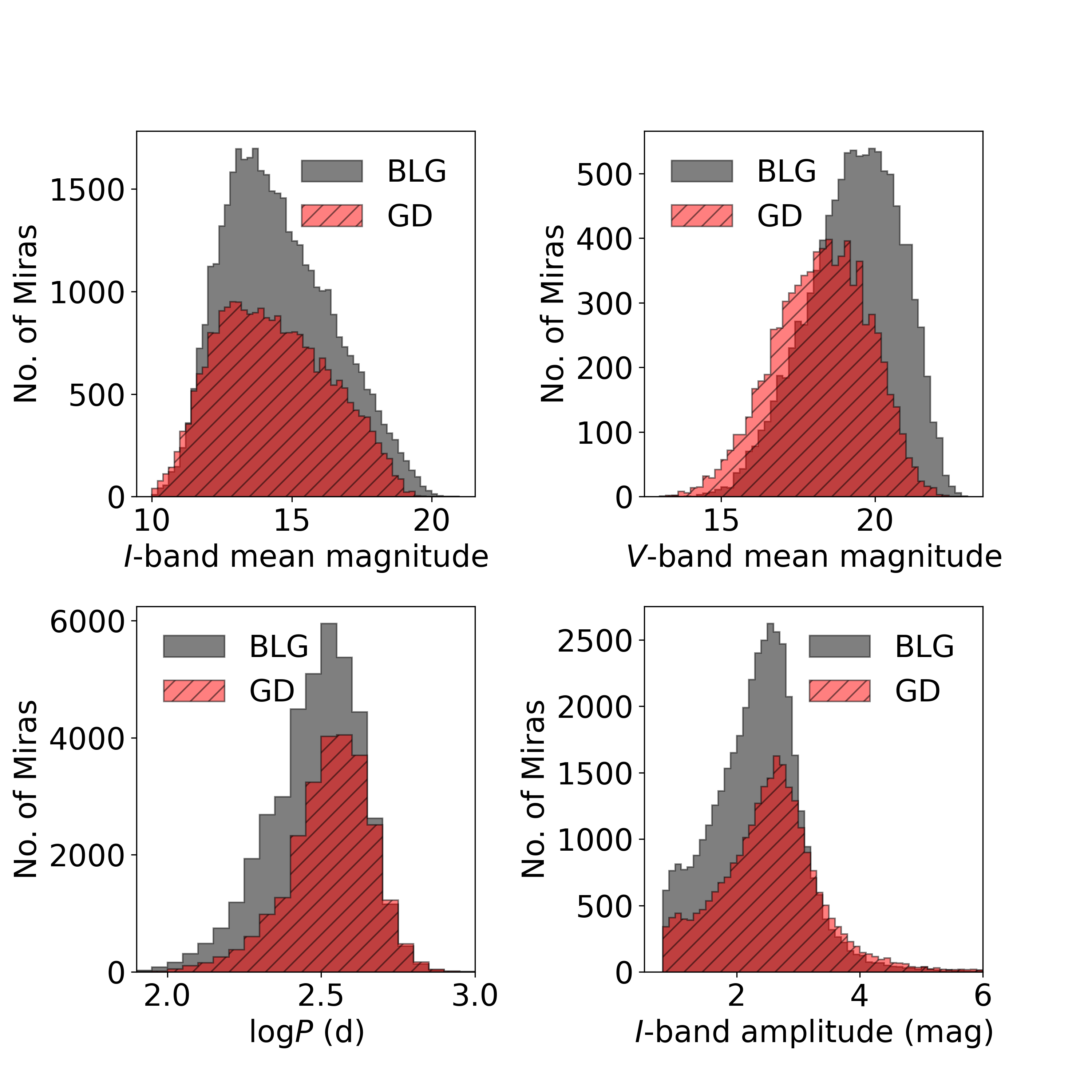}
\caption{Distributions of the {\it I}-band mean magnitudes (top left), {\it V}-band mean magnitudes (top right), pulsation periods (bottom left), and {\it I}-band amplitudes (bottom right) for Miras in the BLG (gray histograms) and GD fields (red dashed histograms).}
\label{fig:properties}
\end{figure}

\subsection{Completeness and purity of the catalog} \label{subsection:completeness}

The completeness of our collection of Mira variables strongly depends on many factors, e.g., the observed brightness of stars, distance to the stars, interstellar extinction toward the stars, stability of light curves, number of epochs, and time-span of observations. Unlike in other types of variable stars, the amplitudes of the light curves do not limit the completeness of the Mira catalogs, since even the minimum amplitude of 0.8 mag (in the {\it I}-band) is much larger than photometric uncertainties of the individual data points. On the other hand, critical factors in the completeness of the catalog of Miras are the number of epochs and time-span of observations, as Miras' pulsation periods are typically longer than 100 days and their light curve shape does not repeat precisely from cycle to cycle. During our search for Mira variables, we found several hundred large-amplitude light curves affected by a small number of epochs and a short time-span for which we could not even estimate their periods. Such objects could be also Miras, however, we did not have enough evidence to prove it, so we rejected such stars from the sample as we did not want to contaminate the catalog.

We are able to estimate the completeness of our collection using Miras observed twice by the OGLE survey, because they were located in the overlapping parts of adjacent fields (for more details see Section \ref{subsection:selection}). Since stars located close to the edges of the fields are usually affected by a smaller number of points, we limited our analysis to the pairs of light curves consisting of at least 100 epochs each. Out of 2630 pairs found, only 1707 pairs met this criterion. Therefore, we had an opportunity to independently detect 3414 counterparts. For 1575 pairs, we successfully identified and classified both Miras, while for the remaining 132 pairs we detected just one Mira star. This gives the completeness of our collection at the level of $96\%$.

We carefully checked all of the 132 light curves that we missed during our selection process. Almost all of them were missed because of the small number of epochs, which made them impossible to be classified as Miras.

The purity of our collection of Mira stars is limited by the LPVs with amplitudes near the boundary between Miras and SRVs (0.8~mag in the {\it I}-band). As described in Section \ref{subsection:reexamined} and presented in Figure \ref{fig:mira_srv_reclassification}, such stars may oscillate between the Mira and SRV phases. On the other hand, the probability of confusing Mira variables with other types of variable stars (different than LPVs) during the visual inspection of their light curves is very low. As a result, we estimate that the purity of our collection is very high, however, it is unfeasible to estimate.

\section{Discussion} \label{section:discussion}

In Figure \ref{fig:properties}, we present histograms of the basic properties of Miras from our collection, i.e., the {\it I}- and {\it V}-band magnitudes, the pulsation period, and {\it I}-band brightness amplitude. We show histograms separately for stars located in the BLG and GD fields.

The distributions of {\it I}-band mean magnitudes (top left panel in Figure \ref{fig:properties}) are limited at the brighter side of the histograms, which is related to the OGLE saturation limits (see Section \ref{section:data}). The large variability amplitudes enable us to detect Miras brighter than the saturation limit, i.e., we have the opportunity to observe them in phases other than maximum light. The same effect is observed for the faintest Miras, which means magnitudes are below the detection limit -- it is possible to observe these stars around their maximum light and classify them as Miras. Approximately two percent of all Miras from our collection may be affected by the proximity of their mean brightness to either the saturation or detection limits. For the brightest, as well as the faintest Miras, their mean magnitudes and brightness amplitudes can be measured with lower accuracy. However, this has the greatest impact for stars with poorly-covered light curves. The maxima of the {\it I}-band magnitude distributions are similar for the BLG and GD populations, at 13.5 mag for BLG, and $\sim 13.0$ mag for GD. The shift of the histogram for the GD population to brighter magnitudes is likely due to the shorter exposure times (so stars saturate at brighter magnitudes), on average shorter distances to these stars, and a lower interstellar extinction.

The spectral energy distributions of Miras show maxima around $1-2$ $\mu$m \citep[see, e.g., Figures 6 and 7 in][]{2021ApJS..257...23I}, therefore the distributions of the {\it V}-band mean magnitude are shifted toward larger values (top right panel in Figure \ref{fig:properties}), with maxima at 19.7 mag for stars in the BLG, and 18.5 mag for Miras in the GD. The distribution of the BLG Miras population is shifted toward fainter magnitudes due to the higher interstellar extinction in this direction and greater depth of the OGLE photometry in the BLG fields compared to the GD ones.

An interesting feature is presented in the distribution of the pulsation periods (bottom left panel in Figure \ref{fig:properties}). The GD Mira population has on average longer pulsation periods than the BLG population, which indicates that GD Miras are on average younger than their BLG counterparts \citep{2022A&A...658L...1T}. Maxima of the pulsation period distributions are at $\log P = 2.52$ ($\sim331$ days) for the BLG population, and at $\log P = 2.56$ ($\sim363$ days) for the GD population. 

To confirm the statistical significance of the difference between the BLG and GD pulsation period distributions, we compared these distributions using the two-sample two-sided Kolmogorov-Smirnov (KS) test with the null hypothesis that both distributions are similar, against the alternative hypothesis that they are different. As the BLG sample is much larger than the GD one, we randomly chose 25,625 stars from the BLG (i.e., same as the GD sample size), and we calculated the KS test. We repeated the sample drawing and calculation of the KS test 1000 times, each time obtaining a p-value close to 0. This means that the null hypothesis could be rejected with high probability, leading to the conclusion that the BLG and GD pulsation periods distributions are significantly different.

In general, the spiral structure of the Milky Way can be traced by molecular gas \citep{2001ApJ...547..792D}, neutral gas \citep{1988gera.book..295B}, star-forming regions \citep{2014ApJ...797...39B, 2014ApJ...783..130R} or young stars, e.g. classical Cepheids \citep{2019Sci...365..478S, 2019AcA....69..305S, 2019NatAs...3..320C}. On the other hand, the second main component of the Milky Way, the bulge, can be traced by red clump stars \citep{1994ApJ...425L..81W, 1997MNRAS.292L..15L, 2010ApJ...721L..28N, 2010ApJ...724.1491M, 2013MNRAS.435.1874W, 2019A&A...627A...3L}, RR Lyrae variables \citep{2015ApJ...811..113P, 2022MNRAS.509.4532S}, or finally Miras \citep{2009MNRAS.399.1709M, 2017ApJ...836..218L, 2020MNRAS.492.3128G}. Recently \citet{2020ApJ...891...50U} suggested, that GD Miras with longer periods probably trace the spiral arms in the Milky Way. It turns out that Mira-type variables, thanks to the wide age range and being abundant throughout the Milky Way, are extremely valuable tracers of old, intermediate, and young stellar populations. 

The distributions of the {\it I}-band brightness amplitude are cut at 0.8 mag, which is related to the classification method of Miras (see Section \ref{subsection:selection}). Miras in the GD fields have on average larger amplitudes with a peak of the distribution at 2.7 mag, while the BLG population has a slightly lower amplitude, with the distribution peak at 2.5 mag. It is clearly seen, that Miras with {\it I}-band amplitudes higher than 4 mag are rare. Additionally, it can be noticed that there is a small surplus of GD Miras with the {\it I}-band brightness amplitudes higher than 3 mag.

Miras are usually divided into O-rich and C-rich \citep[][]{2010ApJ...723.1195R} stars. \citet{2005AcA....55..331S} made such a division for the LMC Miras based on their position in the optical Wesenheit index vs. NIR Wesenheit index diagram. 
Unfortunately, the same division could not be used for LPVs in the Milky Way due to a considerable depth of the BLG and GD along the line of sight \citep{2013AcA....63...21S}. However, our latest research \citep[][Iwanek et al. 2022 in prep.]{2021ApJ...919...99I} shows that such a division can be made based on the pulsation periods and mid-IR color indices.

One of the still unsolved mysteries surrounding the structure of our Galaxy is the existence of the X-shaped bulge. The X-shaped bulge has been originally discovered in the red clump (RC) stars \citep{2010ApJ...721L..28N, 2010ApJ...724.1491M}. However, there are doubts whether the split in the RC stars along some lines of sight is caused by a geometric effect, or it depends solely on the different brightness of the two RC stars populations \citep[e.g., ][]{2019A&A...627A...3L}. The nearly complete collection of the Mira stars, which are also standard candles representing an intermediate-age population, can independently prove or falsify the geometric split of the bulge. \citet{2017ApJ...836..218L} did not find any evidence for the X-shaped bulge in the Mira variables, however, this study was limited to a much smaller sample of Miras than the one presented in this paper. The collection of Miras reported in this paper will be used for a direct examination of the three-dimensional structure of the Galactic bulge.

\section{Conclusions} \label{section:conclusions}
In this paper, we presented and discussed the largest, purest, and most complete collection of Milky Way Miras published to date. In total, we found 65,981 Mira-type variables in the OGLE databases. The vast majority of this collection (47,532 stars) turned out to be new discoveries, and with this catalog, we more than tripled the number of known Miras in the Galaxy.

In the catalog, we provide the J2000 equatorial coordinates, pulsation period, {\it I}-band and {\it V}-band (if available) mean magnitudes, {\it I}-band amplitude, cross-identifications with other variable stars catalogs, and finding chart for each Mira. We also provide the {\it I}-band and {\it V}-band time-series photometry collected since 1997 during the OGLE-II, OGLE-III, and OGLE-IV sky surveys.

This collection will be the basis for many future studies on the stellar evolution or structure of the Milky Way. We believe that our catalog will contribute to solving the puzzle of the X-shape bulge.
Thanks to the high completeness and purity of the catalog, it can also be used for training classification machine learning algorithms aimed at the automatic classification of variable stars in large-scale sky surveys.

\vspace{0.3cm}
We would like to thank the anonymous Referee for constructive and valuable comments that improved our manuscript. This work has been supported by the National Science Centre, Poland, via grants MAESTRO no. 2016/22/A/ST9/00009 to IS and OPUS 2018/31/B/ST9/00334 to SK. PI is partially supported by the {\it Kartezjusz} programme no. POWR.03.02.00-00-I001/16-00 founded by the National Centre for Research and Development, Poland. This research has made use of the International Variable Star Index (VSX) database, operated at AAVSO, Cambridge, MA, USA.

\bibliography{paper}
\bibliographystyle{aasjournal}

\end{document}